\documentclass[namedreferences]{solarphysics}
\usepackage[optionalrh]{spr-sola-addons} 
\usepackage{graphicx}        
\usepackage{color}           
\usepackage{url}             

\newcommand{\apj}{    {\it Astrophys. J.}}
\newcommand{\apjl}{   {\it Astrophys. J. Lett.}}

\newcommand{\mnras}{  {\it Mon. Not. Roy. Astron. Soc.}}
\newcommand{\nat}{    {\it Nature}}
\newcommand{\pasp}{   {\it Pub. Astron. Soc. Pac.}}

\newcommand{\solphys}{{\it Solar Phys.}}

\begin{document}

\begin{article}

\begin{opening}

\title{The Role of Magnetic Fields in Transient Seismic Emission
Driven by Atmospheric Heating in Flares\\}

\author{C.~\surname{Lindsey}$^{1}$\sep
        A.-C.~\surname{Donea}$^{2}$\sep
        J. C.~\surname{Mart\'\i nez Oliveros}$^{3}$\sep
        H. S.~\surname{Hudson}$^{3}$\sep
       }
\runningauthor{Lindsey, Donea, Mart\'\i nez Oliveros \& Hudson}
\runningtitle{Role of Inclined Magnetic Fields in Transient Seismic Emission
due to Atmospheric Heating}

   \institute{$^{1}$ NorthWest Research Associates
                     email: \url{clindsey@cora.nwra.com} \\ 
              $^{2}$ Monash University, Melbourne, Australia
                     email: \url{Alina.Donea@monash.edu.au} \\
              $^{3}$ Space Sciences Laboratory, 
University of California at Berkeley, Berkeley, California \\
                     email: \url{Oliveros@ssl.berkeley.edu} \\
             }

\begin{abstract}
Fifteen years after its discovery, the physics of transient seismic emission
in flares remains largely mysterious.
An early hypothesis by its discoverers proposed that these ``sunquakes''
are the signature of a shock driven by ``thick-target heating'' of the 
flaring chromosphere.
H-$\alpha$ observations show evidence for such a shock in the low
chromosphere during a flare.
However, simulations of shocks driven by impulsive chromospheric heating
show withering radiative losses as the shock proceeds downward.
The compression of the shocked gas not only heats it but increases its
density, making it more radiative.
So, radiative losses increase radically with the strength of the shock. 
This has introduced doubt that sufficient energy from such a shock
can penetrate into the solar interior to match that indicated by the
helioseismic signatures.
The purpose of this paper is to point out that simulations of acoustic
transients driven by impulsive heating have yet to account for magnetic
fields characteristic of transient-seismic-source environments.
We show this to be a critically important factor that has a major impact
on the seismic flux conducted into the solar interior.
A strong horizontal magnetic field, for example, greatly increases the 
compressional modulus of the chromospheric medium.
This greatly reduces compression of the gas, hence the radiative losses 
as the transient passes through it.
This could explain the strong affinity of seismic sources to regions of
strong, highly inclined penumbral magnetic fields, including the neutral
lines separating opposing polarities in $\delta$-configuration sunspots.
The basic point, then, is that {\it the role of inclined magnetic fields
is fundamental to our understanding of the role of impulsive heating
in transient seismic emission}.
Obliquely inclined magnetic fields will complicate simulations of
impulsive heating considerably.
However, horizontal magnetic fields, as a preliminary control simulation,
can be incorporated into standard 1-D thick-target-heating simulations
with a relatively simple adaptation of existing HD codes. 
\end{abstract}
\end{opening}

\section{Introduction}
     \label{S-Introduction} 

The first known instance of seismic transient emission, also known as a
``sunquake,'' was that discovered by \inlinecite{Kosovichev98}, emanating
from the X2.6-class flare of 1996 July 9 in the declining phase of 
Cycle 22.
Seven years passed, with some major acoustically quiet flares, before 
\inlinecite{Donea05} found seismic emission from multiple sources
in the flares of 2003 October 28--29.
This led to the subsequent discovery of about a dozen instances of
transient seismic emission in the declining phase of Cycle 23, some 
associated with relatively weak flares.
Transient seismic emission is now known to be a moderately common
phenomenon.
Nevertheless, something about the mechanics behind this phenomenon
continues to elude us.

The significant manifestation of transient seismic emission is a
pattern of outgoing ripples seen in Doppler-helioseismic observations
15--45~min following the impulsive phase of a flare.
Computational seismic holography applied by \inlinecite{Donea99},
\inlinecite{Donea05} and others, showed compact sources up to
$\sim$12~Mm across at sites of hard X-ray emission, and sometimes
$\gamma$-ray emission.
Estimates of the seismic energy, $E_S$, emitted by a single compact source,
based upon signatures derived from seismic holography
\cite{Donea99,Donea05}, are frequently in the range
10$^{27}$--10$^{28}$~erg.%
\footnote{Computational seismic holography is based upon the extrapolation
of the acoustic field from a surface region several thousand km from a
source region back to the source region itself.
Estimates of the energy flux from this signature are the acoustic analogy
of photometry in the electromagnetic spectrum.
For an example of this application, we refer to \inlinecite{A-G12}.}
The vertical momentum carried by a seismic transient at the solar surface
is estimated at $E_S/c$, where $c$ is the sound speed.
Taking $c$ to be $\sim$10~km~s$^{-1}$ renders seismic momenta in the
range 10$^{22}$--10$^{23}$~dyne~s.

In large flares, there may be more than one such source, in which case
the total seismic energy emitted can be correspondingly greater.
This acoustic energy is only a small fraction of the total energy 
\cite{Emslie12} estimated for a flare, generally a fraction of a percent,
and for large flares often less than 10$^{-4}$ of the total.

Attempts to understand transient seismic emission have focused upon
three mechanisms:

\subsection{Impulsive Heating of the Chromosphere} 
  \label{S-TTH}

\inlinecite{Kosovichev98} attributed transient seismic emission
to a pressure wave driven by ``thick-target heating'' of the chromosphere
by energetic electrons.
A good deal of what we understood of this mechanism at the time came out
of 1-D numerical HD simulations (Fisher, Canfield and McClymont 1985a,b,c,
thenceforth FCM) applied to pre-flare models of plage chromospheres and
coronae.
Energetic electrons accelerated in the corona heat the
upper chromosphere to something like coronal temperatures.
The initial pressure of the medium, 10--100~dyne~cm$^{-2}$ is increased
by about two orders of magnitude.
The resulting increase in pressure drives a shock downward into the medium 
beneath the heated layer, carrying a momentum of the order of
$10^{22}$~erg~cm~s$^{-1}$ (\opencite{Kosovichev95}).
This is essentially balanced by the upward momentum of the heated layer,
which explodes upward into the pre-flare corona.
\inlinecite{Zarro89} found this shock model consistent with H-$\alpha$
emission spectra typical of M-class flares.

More recent consideration to the role of impulsive chromospheric
heating in transient seismic emission is given by \inlinecite{Zharkova08}
and \inlinecite{Zharkov11}.

The hypothesis that transient seismic emission is driven by impulsive 
chromospheric heating has the problem that severe radiative losses
deplete the transient as it passes through the underlying low
chromosphere.
These concerns were first communicated to us by Fisher (2006, private
communication).
They are mentioned by \inlinecite{Hudson08} and treated in further
detail by \inlinecite{Fisher12}.
This conclusion is reinforced by FCM-style simulations by 
\inlinecite{Allred05}, and, with improved treatment of radiative
transfer, by Allred (private communication, 2012).
Even in acoustically active flares the signature of heating is 
often evident in regions from which no seismic emission appears to emanate.
This includes acoustically inactive flares, from which no transient
seismic emission at all is evident.

Our understanding of the dynamics and geometry of atmospheric heating
has changed dramatically since FCM.
The magnetic structure of penumbrae, from which transient seismic emission
generally appears to emanate, is highly filamentary, as opposed to the 
plane-parallel models of FCM.
\inlinecite{Fletcher08} suggest that the high-energy electrons that serve
as the primary agent for the evident heating could be accelerated much
deeper than in the models of FCM, i.e., in the chromosphere instead of
the corona.
Indeed, recent {\it SDO}/STEREO/RHSSI observations by \inlinecite{Martinez12}
using parallax to determine the source heights of white light (WL) and
hard X-ray (HXR) emission associated with atmospheric heating indicate
that this heating is significantly deeper than in the models of FCM.

\subsection{Backwarming of the Photosphere} 
\label{S-BW}
    
Based upon some of these developments, \inlinecite{Donea99} and 
\inlinecite{Moradi07} proposed that heating of the photosphere by visible
and near-UV continuum radiation from the heated chromosphere could drive
a seismic transient (see also \opencite{Lindsey08}).
In this model, a pressure transient is launched at the base of the 
photosphere, which most efficiently absorbs the continuum radiation
from the chromosphere, and proceeds downward (and upward) from thence.
Once the downward component of the transient penetrates into the solar
interior, further radiative losses are blocked by the high opacity of
ionized hydrogen.
This mechanism, known as ``backwarming,'' was suggested by sources of 
seismic transient emission that coincided closely with transient 
white-light flare kernels (\opencite{Lindsey08}).

Instances have since been discovered of strong seismic emission from
regions that show no appreciable excess in visible continuum radiation
integrated over the neighborhood of the seismic source.
An example is the X2.2-class flare on 2011 February 15.
Integrated over a circular disk whose area is 90~Mm$^2$, encompassing the
seismic source, these fluctuations are only 0.12\% of the quiet-Sun
intensity.
For comparison, a similar average taken over the seismic source of the
white-light flare of 2003 October 29 found a continuum emission 10\%
in excess of the quiet-Sun intensity \cite{Donea05}, i.e., about 80 times
greater than in the flare SOL2011-02-15 \cite{A-G12}.
According to our rough understanding of how back-warming could
prospectively work as a seismic source \cite{Lindsey08}, a compression
wave driven by heating commensurate with this degree of fluctuation in
continuum intensity could explain barely 2\% of the the seismic energy
that appears to emanate from the flare SOL2011-02-15.
Because of this, it is evident that some other mechanism must be in
operation to explain the seismic transient released by this flare.

Variations to the back-warming theme draw upon energetic protons 
\cite{Zharkova08}.
Protons with energies of 200~MeV could penetrate to the base of the
photosphere, heating it directly.
Protons of these energies interact with the nuclei of elements such as
oxygen, iron and magnesium \cite{Najita70} to excite $\gamma$ rays.
Such $\gamma$ rays are indeed observed in some seismically active flares.
However, we now know of instances (\opencite{Donea06}) of strong seismic
emission from flares from which no $\gamma$ radiation was detected.

\subsection{Lorentz-Force Transients} 
\label{S-LFT}
    
Based largely upon the foregoing assessments, \inlinecite{Hudson08} and
\inlinecite{Fisher12} proposed that transient seismic emission is the
result of Lorentz-force transients, caused by the sudden re-configuration
of the magnetic field in the impulsive phase of the flare.

Magnetic signatures roughly consistent with Lorentz-force transients
driving transient seismic emission consistent with helioseismic signatures
are reported by \inlinecite{Donea06}, \inlinecite{Sudol05}, 
\inlinecite{Wang02} \inlinecite{Wang10}, \inlinecite{Petrie10}, 
\inlinecite{Petrie12} and others.
These signatures appear to be encumbered by two significant qualifications:

\begin{itemize}
\item Magnetic signatures representing highly perturbed radiative
environments such as are typical of flares are notoriously unreliable,
often showing momentary reversals in the field direction inconsistent
with expectations based upon MHD in a highly conducting medium.

\item Such magnetic signatures are common in acoustically inactive flares,
as they are in acoustically active flares in wide-spread regions from
which no detectable seismic emission emanates.
\end{itemize}

\inlinecite{A-G12} studied the seismically active flare SOL2011-02-15,
concentrating on the magnetic signature in the source region of the strongest
of two sources (Figure~1).
Based upon line-of-sight magnetic signatures immediately before the impulsive
phase and 10--15 minutes after, they offer a somewhat discouraging appraisal
of the potential of the local Lorentz force transient as a driver for the
seismic emission observed.
However, they do not rule it out.
Further consideration to the role of Lorentz-force transients in this flare
is given by \inlinecite{Zharkov11}, based largely upon the apparent magnetic
connectivity between the primary acoustic source and a second one, which they
themselves discovered.

\begin{figure}    
\centerline{\includegraphics[width=1.0\textwidth,clip=]{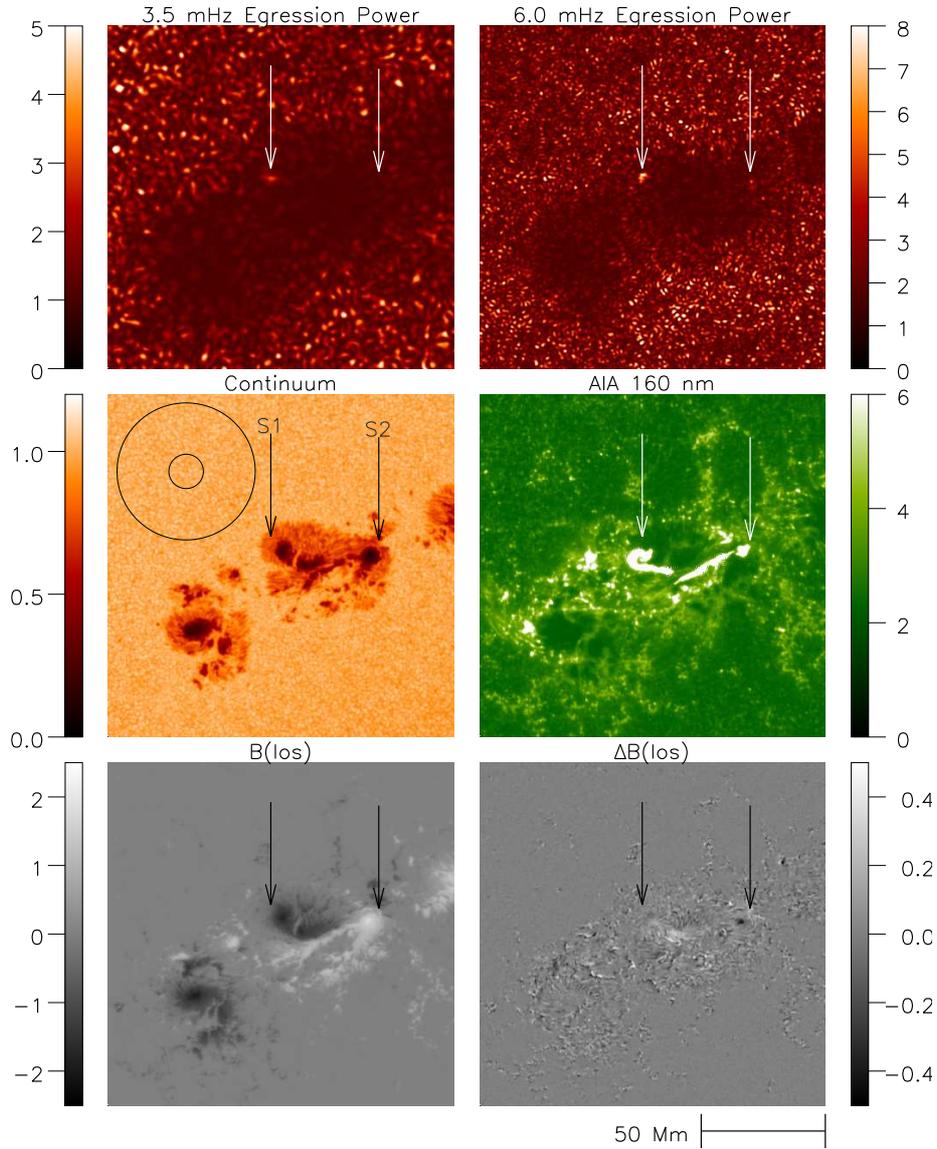}}
\caption{Doppler egression-power maps from {\it SDO}/HMI
observations representing 1-mHz bandpasses centered at 3~mHz (upper left)
and 6~mHz (upper right) in units of m$^2$~s$^{-2}$.
Middle row shows a pre-flare HMI continuum intensity snapshot (left)
and an AIA snapshot (right) of 160~nm continuum radiation in the early
impulsive phase, both normalized to unity for the quiet Sun.
Bottom row shows a pre-flare HMI line-of-sight magnetogram (left), and
line-of-sight difference in the magnetic field from before to after the
impulsive phase, the time difference being 15 min, both in units of
kGauss.
Arrows locate compact sources of seismic emission in the upper right
egression-power map.
Concentric circles in the upper left frame show the dimensions of the
pupil applied in the helioseismic egression computations in the top row.}
\label{F-eg-pwr-maps}
\end{figure}

AIA observations show strong 160~nm emission from the end
of a ribbon $\sim$2.8~Mm SSE of the seismic source in the early impulsive
phase (middle-right panel of Figure~1).
Within seconds, emission in this region is obscured by saturation/overflow
from strong, extended emission at large.
{\it Hinode}/SOT observations in the CaII K-line show a very similar
ribbon that sweeps northward through the acoustic source region
in the minute or so following the time of the 160~nm snapshot.
This suggests strong heating of the chromosphere, that might drive a 
seismic transient---{\it if} the resulting shock could somehow escape 
the radiative losses described in \S \ref{S-TTH}, above.
\inlinecite{A-G12} report HXR emission whose RHESSI signature encompasses
the seismic source region, which would connect the heat source to
energetic electrons.\footnote{The spatial resolution of the HXR 
observations, at $\sim$1.7~Mm.}
Like the magnetic signatures, both the 160-nm and the HXR sources are
stronger in other far-removed regions, from which no seismic 
emission is detected.

\section{Regional Selectivity}
     \label{S-Selectivity} 

In summary, then, the signatures indicating impulsive chromospheric heating
and Lorentz-force transients appear to share one conspicuous quality in
common, both in the flare SOL2011-02-15 and in other flares,
acoustically active or not:
They all show strong, apparently transient signatures in vast regions
from which no significant seismic emission emanates.
It appears that there has to be something special about seismic-source
regions such that any of the mechanisms yet considered (or perhaps others)
could have a role in transient seismic emission.
For either chromospheric heating or Lorentz-force transients to satisfy
the helioseismic signatures, something distinctive about their respective
operations begs to be identified that makes it {\it horizontally selective}
of the highly restrictive regions from which seismic radiation emanates.
This is the primary theme of this study.
Failing such an identification, it should begin to appear that seismic
transient emission must depend upon some mechanism not yet considered.

In the case of Lorentz-force transients, for example, a basis for
``regional selectivity'' might be the difficulty of credible magnetic
measurements during the impulsive phases of flares.
If transient magnetic signatures during the impulsive phase of the flare
are simply spurious, then magnetic variations that actually occur in
the impulsive phase may be insufficiently sudden to excite a seismic
transient---except where the signatures of transient seismic sources
actually appear.
Magnetography is known to be highly susceptible to this liability under
flaring conditions.
If this is the basis of the regional selectivity we find, then what
is distinctive about seismic source regions so that magnetic transients
in them are more sudden than elsewhere?

For transients driven by impulsive heating, a highly suggestive basis
for regional selectivity is the impact of a strong magnetic field on the 
overbearing radiative losses discussed in \S 1.1, which discourage the
downward conduction of seismic transients in non-magnetic simulations.
In the absence of a magnetic field, a shock driven by the expansion of
the heated medium compresses the underlying medium, heating it and
increasing its density.
Both the heating and the increased density make the gas a more effective
radiator.
Hence, the radiative losses increase strongly, nonlinearly, with the
degree of compression, which is 1--2 orders of magnitude in the low
chromosphere in the simulations of Fisher, Canfield \& McClymont
(1985a,b,c).
The onus appears to be whether any amount of impulsive heating can
manifest a seismic transient that can survive the radiative losses
entailed in passage through both the low chromosphere and photosphere
to penetrate into the solar interior.

As a screening measure to judge whether impulsive heating merits
serious consideration, simulations such as those of \inlinecite{Fisher12}
and \inlinecite{Allred05} can seem discouraging indeed.
However, we want to point out just the opposite, that {\it these simulations
are entirely consistent with helioseismic signatures as we know them.
Impulsive chromospheric heating simulations to date have been computed only
for \underline{non-magnetic} media, from which detectable transient
seismic emission has indeed never emanated, to our knowledge.}
Based upon this, we propose, as a hypothesis, that the very radiative losses
that discourage the impulsive heating as a driver of transient seismic
emission in the context of a non-magnetic medium are the key to the high
degree of regional selectivity shown by transient seismic emission in 
flares.
Essentially all transient seismic sources appear in sunspot penumbrae,
regions known to be infused with strong, highly inclined magnetic fields.
We want to propose that a major mechanical role of this field can be
a {\it radical reduction in radiative losses} in those localities.
This reduction in radiative losses could rescue impulsive heating
as a tenable source of the seismic emission that emanates from some
flares.
Moreover, if backed up by appropriate MHD simulations that incorporate
oblique magnetic fields, and by appropriate observational diagnostics,
it may very well explain the high regional selectivity that characterizes
transient seismic sources.

\section{The Mechanical Role of an Inclined Magnetic Field in 
Seismic-Transient Conduction}
\label{S-mechanics}      

For a rough, preliminary understanding of the role of an inclined field,
we consider the simplest case, that of a horizontally uniform medium
infused with a horizontal field, likewise horizontally uniform, into
which we introduce a horizontally uniform shock, driven by a horizontally
uniform thick-target-profile heating.
In the MHD context the action of the horizontal magnetic field is 
essentially equivalent to that of a large increase in the compressional
modulus of the medium, the magnetic component of which is
\begin{equation}
\kappa_M ~=~ -V{\partial p_M \over \partial V} ~=~ 2p_M,
\label{eq-mag-mod}
\end{equation}
where 
\begin{equation}
p_M \equiv {B^2 \over 8\pi}
\label{eq-mag-pressure}
\end{equation}
is the magnetic pressure, with $B$ the magnetic field strength, and $V$
represents the volume of the local unit of medium being vertically 
compressed by the shock.
The compression, then, is perpendicular to $\bf B$, the local magnetic
induction.
For a magnetic field of 500~Gauss%
\footnote{This exercise selects the lower limit of our understanding
of the typical range of the horizontal components of penumbral fields
as an example.
\cite{Borrero11} find penumbral fields with horizontal components in the
range 1000--1200~Gauss, with a stochastic scatter of $\pm500$~Gauss.} 
in the primary source region of the flare
SOLA2011-02-15, this is $\sim$40 times the isothermal gas modulus, 
\begin{equation}
\kappa_g ~=~ -V\Big({\partial p_g \over \partial V}\Big)_T ~=~ p_g,
\label{eq-gas-mod}
\end{equation}
where $p_g$ is the local gas pressure.


The presence of a horizontal magnetic field whose strength is typical of
sunspot penumbrae, then, will make the medium much more ``vertically 
rigid,'' i.e., much more resistant to compression perpendicular to the
field.
The medium will therefore conduct seismic energy downward with a much smaller
degree of compression.
The radiative losses will be reduced accordingly.
Since the radiative losses in the non-magnetic medium increase strongly,
i.e., highly non-linearly, with compression, we have to suspect that the 
radiative losses will be a small fraction of those indicated by the
non-magnetic simulations, for a sufficiently inclined field.
This will introduce a high degree of selectivity towards regions of highly
inclined magnetic field.
In the ideal case of a horizontal field conducting a downwardly propagating
transient, nearly all of the compressional work will be done on the magnetic
field, which is almost completely elastic, hence lossless.

The role of inclined magnetic fields, then, needs to be seriously
considered as a prospective explanation of strong regional selectivity
of transient seismic sources in favor of strongly inclined magnetic fields.
Indeed, among the most conspicuous such source regions is the penumbral
neutral line separating umbrae of opposing polarities in $\delta$-configuration
sunspots.
In these instances, e.g., \inlinecite{Moradi07} (see Figure 2 of this study),
the magnetic field is exceptionally strong, leading to an especially high
magnetic modulus, $k_M$, and is essentially horizontal, hence avoiding 
losses due to coupling between fast and slow modes (see \opencite{Cally11}).

\begin{figure}    
\centerline{\includegraphics[width=1.0\textwidth,clip=]{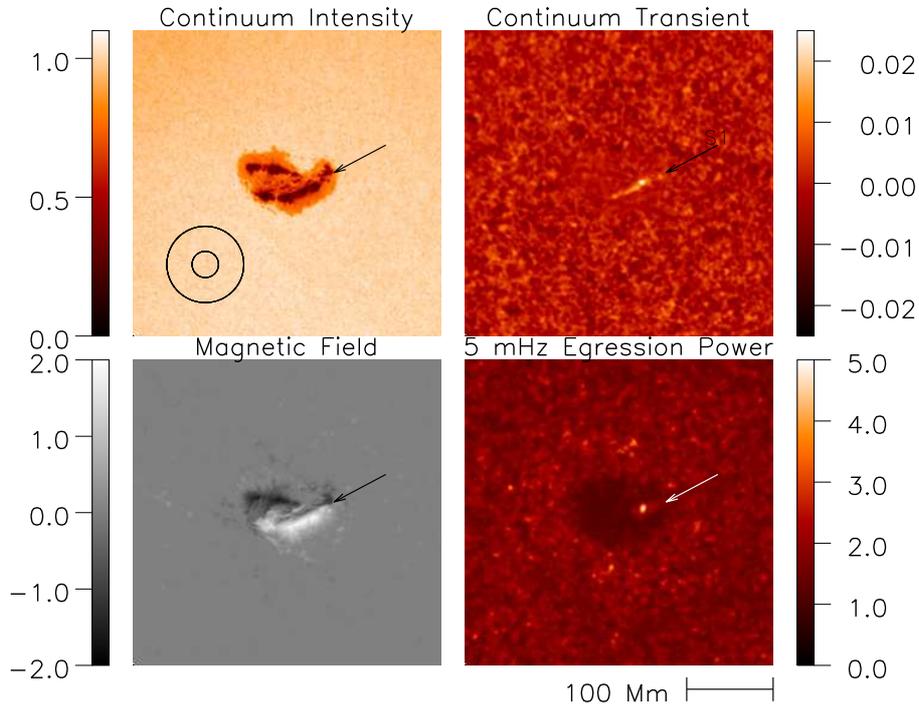}}
\caption{Doppler egression-power map (lower right) of seismic emission 
from NOAA AR10720 at 00:42~UT, during the impulsive phase of the flare of
2005 January 15.
The line-of-sight magnetogram (lower left) identifies AR10720 as a
$\delta$-configuration sunspot.
Upper left frame shows the pre-flare continuum intensity, from
{\it SOHO}/MDI.
Upper right frame shows the continuum-intensity excess in the impulsive
phase of the flare, from GONG.
Bottom left frame shows a line-of-sight magnetogram shortly before flare
onset.
Concentric circles in the upper left frame show the dimensions of the
pupil applied in the helioseismic egression computation.}
\label{fig-delta-spot}
\end{figure}

\section{Qualifications}
      \label{S-qualifications}      

\subsection{Magnetic Encumbrance of the Work Capacity of a Heated Gas}
      \label{S-mag-encumberance}      

Somewhat anticipating realistic simulations, it should be noted that
the magnetic rigidity that should (we propose) so greatly reduce radiative
losses once a transient has development must also inhibit this development.
This is because it encumbers the expansion of the heated medium, which is
the agent of mechanical work upon its surroundings.
Heating applied to a low-$\beta$, i.e. magnetically rigid, medium will
respond with less expansion than a non-magnetic medium.
At some point this will limit the capacity of the heated gas to do work
on its surroundings, including on the magnetic field itself, hence the
efficiency with which impulsive heating will drive a seismic transient.
For an estimate of this encumbrance over a range of temperatures to which
the gas can be heated, we determine the ``mechanical work capacity'' with
and without a magnetic field for a unit mass (1 gram) of hydrogen, with a
helium abundance of 10\%, initially at temperature $T_0$ of 4,000~K
and under a gas pressure, $p_{g0}$, of 500~dyne~cm$^{-2}$.
The gas is first heated at constant volume, $V_0$, to a temperature 
$T_0 ~+~ \Delta T$, 
raising its pressure as illustrated by the vertical displacements in the
$p$-$V$ trajectories plotted in Figure~3. 
In this representation, the non-magnetic gas is represented by the blue
locus.
The green locus represents an identical gas with a horizontal magnetic 
field initially of 500~Gauss.
Once heated, these gas samples are expanded vertically and adiabatically to
their original pressures, and subsequently allowed to cool at the original
pressure to the original temperature of 4,000~K.
We recognize the areas of the respective closed curves in the $p$-$V$ plane
as the respective adiabatic work capacities,
\begin{equation}
\Delta W ~=~ \oint_{\cal R} p dV,
\label{work}
\end{equation}
of the heated gas samples.

\begin{figure}    
\centerline{\includegraphics[width=1.0\textwidth,clip=]{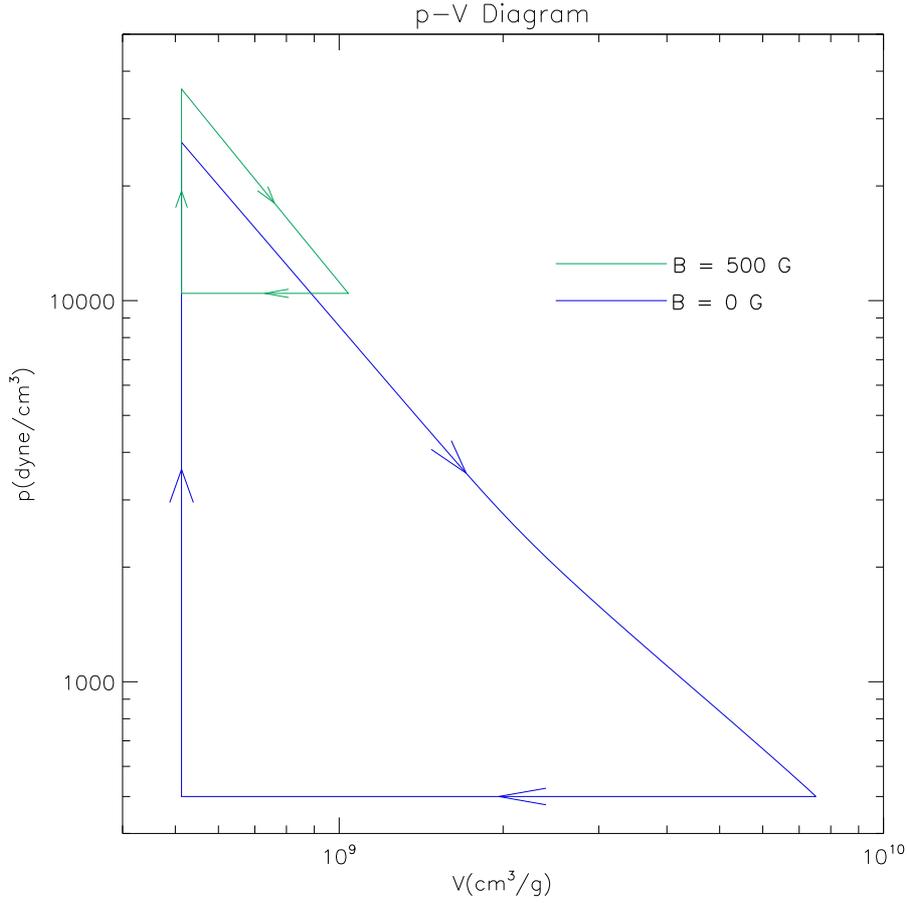}}
\caption{Trajectories in $p$--$V$ space of a unit mass (1~g) of chromospheric
hydrogen-helium initially at a temperature $T$ of 4,000~K and
pressure of 500~dyne~cm$^{-3}$ heated to 10$^5$~K, then expanded 
adiabatically to the initial pre-heated pressure, and finally cooled at
that pressure.
The blue locus represents a non-magnetic medium.
The green locus represents the same cycle for a medium infused with a
500-Gauss magnetic field.
We recognize the respective ``adiabatic work capacities,'' $\Delta W$, 
of these cycle to be the area in $p$--$V$-space enclosed by the respective
curves.
For the foregoing parameters, $\Delta W$ is 
1.44$\times$10$^{13}$~erg g$^{-1}$ for the non-magnetic medium and
0.459$\times$10$^{13}$~erg g$^{-1}$ the magnetic medium.}
\end{figure}

\begin{figure}    
\centerline{\includegraphics[width=1.0\textwidth,clip=]{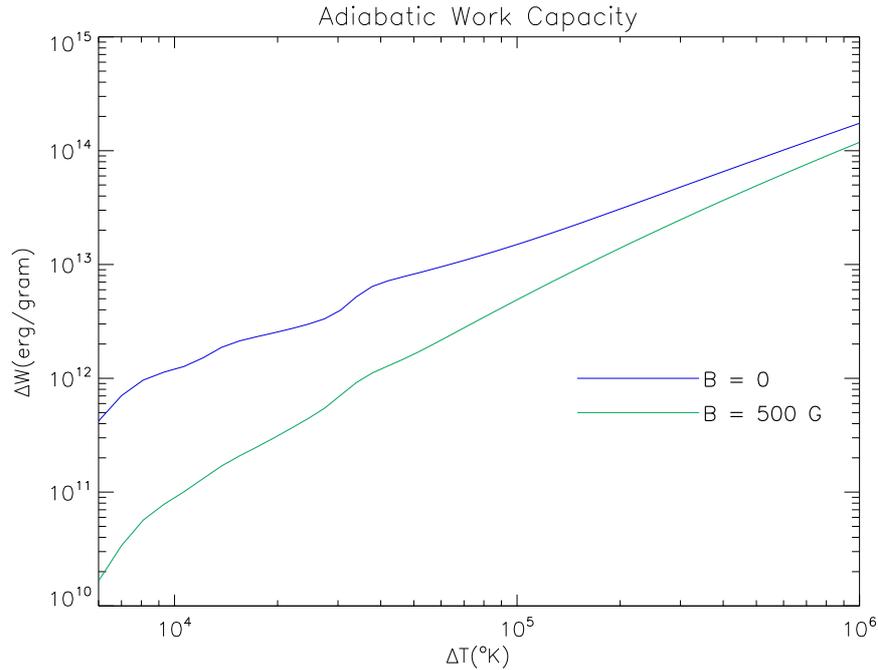}}
\caption{Mechanical work capacity of an unmagnetized heated gas (blue)
is compared with the same for a gas infused with a horizontal magnetic
field whose initial strength is 500~Gauss is plotted as a function of 
the temperature increase, $\Delta T = T_1 ~-~ T_0$, in the heated gas.
The gas, in thermal equilibrium with a helium abundance of 10\%, is 
heated at constant volume from an initial temperature of 4,000~K
to the temperature indicated by the abscissa.
It is then expanded adiabatically until the composite of its gas and magnetic
pressures has returned to the pre-heated composite.
Ordinate values indicate the work done by the expanding gas and magnetic
flux to occupy a greater volume than before the gas was heated.}
\end{figure}

Figure~4 shows comparative plots of $\Delta W$ for temperature increments 
up to $10^6$~K.
The capacity of a magnetized layer to do mechanical work on its surroundings 
is seen to roughly approach that of the non-magnetized gas when the heating
is sufficient to increase the $\beta$ of the gas to something like unity or
more.  
I.e., nowithstanding that the magnetic modulus, $\kappa_M$, strongly
dominated gas dynamics in the low chromosphere in pre-flaring conditions,
if the heating is sufficient to elevate the gas modulus, $\kappa_g$, so that
it dominates over the magnetic modulus, then the impact of the magnetic field
upon the work capacity of the gas becomes relatively modest.
If heated to coronal temperatures, i.e., $10^6$~K, a 500-Gauss magnetic
field reduces the work capacity by about a third.

A magnetic medium, then, is considerably more conducive to
seismic-transient generation if the medium heated is relatively dense.
This suggests that higher-energy particles---which, in a thick-target
model, for example, can penetrate to greater column densities---might
be more conducive to seismic transients than low-energy particles, which
would expend their heat in low-density media.
This might explain the strong association of seismic transients with
hard X-ray sources.

\subsection{Regional Selectivity of Lorentz-Force Transients}
      \label{S-lforce-selectivity}      

It should also be noted that seismic emission due to Lorentz-force
transients, could be fairly selective in favor of regions of inclined
magnetic field---under conditions that are similarly subject to
observational control.
This could be the result of magnetic measurements sampling a layer not
very far above that of ``unity $\beta$,'' beneath which the gas pressure,
$p_g$, dominates the magnetic pressure, $p_M$, and, due to the high
conductivity of the solar plasma, the magnetic field is rigidly frozen
into it.
HMI magnetic diagnostics, for example, are based upon the line
Fe~I~6170~\AA , whose core forms about 100~km above the height at
which $\tau(5000~\AA )$ passes through unity.
The gas pressure, $p_g$, at this height is about a third of the
magnetic pressure, $p_M$, for a 500-Gauss field.
Proceeding downward, from 140~km beneath this level, the gas pressure
dominates, $e$-folding about every 100~km.%
\footnote{The foregoing parameters actually describe the ``standard 
(quiet-Sun) model'' of \inlinecite{JCD93}.
For the sunspot penumbra, the transition will have to be sharper, because
of the Wilson-depressed photosphere of the sunspot \cite{Lindsey10}.}
The horizontal scale of seismic sources such as that labeled ``S1'' in
Figure~\ref{F-eg-pwr-maps} is $\sim$2.5~Mm, about 20 times the $e$-folding
depth in pressure.

We will now consider the simple model in which the magnetic flux we observe
passes from the surface in which it is measured into a rigidly frozen-in
condition in an infinitesimally thin layer beneath it, calling this the
``thin-unity-$\beta$-layer approximation,'' or simply the ``thin-layer
approximation'' for short (see \opencite{A-G12}).
For simplicity, we represent Lorentz forces in the plane-parallel
approximation adopted by \inlinecite{Hudson08}.
The local transient pressure exerted upon the photosphere by a magnetic
transient, $\delta {\bf B}$, localized thereat can be expressed by
\begin{equation}
\delta p_M ~=~ {1 \over 8\pi}(2B_z\delta B_z 
             ~-~ {\bf B}_h\cdot \delta{\bf B}_h).
\label{lforce-xient}
\end{equation}
In the thin-layer approximation, the surface flux density, i.e.,
$B_z$, is held constant during the flare, since the flux is rigidly frozen
into the dense, highly conducting medium just beneath where it is observed.
Hence, $\delta B_z$ in equation (\ref{lforce-xient}) must be null, whence
$$\delta F_z ~=~ -{1 \over 8\pi} {\bf B}_h\cdot \delta{\bf B}_h.
\eqno{(4)}$$

Under the thin-layer approximation, then, seismic transient excitation 
due to differential Lorentz-force transients, $\delta {\bf B}_h$, would
vanish when $\bf B$ is vertical, lending a distinct preference to regions
whose magnetic fields have a strong horizontal component.

Whether the thin-layer approximation is a realistic representation of MHD 
in the environments of seismic sources to a tenable precision on the 
impulsive-phase time scales required for efficient transient seismic
emission has not yet begun to be realistically addressed to our knowledge.
The point of the foregoing exercise, then, is not to promote the accuracy
of the thin-layer appoximation, nor to encumber Lorentz-force-driven
seismic transients with any of a range of qualifications that would follow
from it.
It is only to render a recognition of how {\it some} degree of regional
selectivity {\it could} apply to seismic sources driven by Lorentz-force
transients as we propose they do to impulsive chromospheric heating.
The more important point is that we have abundant resources with which
to address this question in lieu of having the answer to it now.

\section{What Should We Do?}
      \label{S-should-do}      

Considerable attention is now being paid to Lorentz-force transients
as a prospective source of transient seismic emission, motivated by
\inlinecite{Hudson08} and \inlinecite{Fisher12}.
This is only appropriate.
If Lorentz-force transients contribute significantly to transient
seismic emission, the implications respecting the flare mechanism
at large have to be formidable.
If not, the implications of ``not'' are probably similarly formidable.
This might {\it suggest} an onset mechanism that, at least on the time
scale of seismic transients, delivers a great deal of energy to a very
small mass, i.e., high energy electrons, but, as \inlinecite{Zarro89}
suggest, need not impart anything like the $10^{22}$-dyne-s impulse
required by a seismic wave.
The primary goal of this study is to establish the need for a
commensurate expenditure of attention on the contribution of impulsive
chromospheric heating, particularly the need to include inclined magnetic
fields in simulations thereof.
An appreciation of this distinction, if it applies, could lend us some
very cogent clues that we have yet to secure about how flares work.


\section{How Do We Do It?}
\label{S-how-to-do-it}      

Even 1-D simulations in a medium with an obliquely inclined field will
require a considerable extension of standard non-magnetic 1-D simulations
of impulsive-heating dynamics.
The Lorentz force will likewise be oblique, introducing a horizontal
component of motion.
This will introduce the complication of coupling between modes whose
restoring force is primarily compressional and modes whose restoring
force is primarily magnetic tension (\opencite{Cally11}), the latter
of which will add to the depletion of the former, hence of the 
eventual helioseismic signature---thereby further contributing to the
regional selectivity attendant to inclined fields.
The horizontal motion will break azimuthal symmetry, complicating the
RT computations.  
In particular, the relative Doppler-shifts in the chromospheric lines
will depend upon the azimuth of the direction of the radiation.
So, line intensities, opacities, source functions, etc., will require
an account for azimuthal dependencies.
Realistic 3-D simulations will entail further complications still,
and a greater expense in computing resources.
However, for the special case of 1-D simulations of a downwardly
propagating shock in a {\it horizontally} magnetized medium, i.e., a
perpendicular wave or shock, azimuthal symmetry is restored, and the
foregoing complications essentially vanish.
This problem can be addressed by incorporating the magnetic modulus
expressed by equation (1) into the the gas dynamics as if this is simply
a thermodynamic property of the gas.

A feasible approach to the issue of selectivity, then, would be
adaptations of existing HD algorithms applied to simulations such as
those applied by \inlinecite{Fisher12} and \inlinecite{Allred05} to a 
non-magnetic medium to represent a horizontally magnetized medium by
the inclusion of an appropriate elastic modulus (equation 1) in addition
to the thermally induced forces.
If these computations indicate the survival of transient seismic emission
penetrating into the solar interior consistent with helioseismic
signatures from sunspot penumbrae, this can serve as a practical basis
for the investment needed for simulations of chromospheres in which the
magnetic inclination is other than horizontal, i.e., oblique.
These simulations will be needed for an assessment of the degree
of regional selectivity that realistically characterizes
impulsive chromospheric heating as a contributor to transient seismic
emission as indicated by helioseismic signatures.
At the very least, it is important to recognize that the relatively
simple extension of the 1-D simulations of \inlinecite{Fisher12}
\inlinecite{Allred05} to include horizontal magnetic fields is very
timely at this juncture.
Given the strongly selective character of transient seismic emission in
regions of strongly inclined magnetic field, impulsive chromospheric
heating as a driver of transient seismic emission should not be ruled
out until at least this adaptation has been made and applied at length.

\section{Summary}
      \label{S-Summary}      

Simulations of impulsive chromospheric heating as a driver of transient
seismic emission from flares show strong chromospheric shocks with
energies roughly consistent with helioseismic observations of this
phenomenon.
However, these simulations also show radiative losses that deplete 
nearly all of the energy initially investmented into the shock as
it plows downward into the underlying chromosphere and photosphere. 
This has introduced doubt that sufficient energy from such a shock
can penetrate into the solar interior to match that implied by the
helioseismic signatures.
The inclusion of a highly inclined magnetic field, typical of
transient-seismic-source regions would have a major impact upon 
this assessment.
The magnetic field can greatly increase the compressional modulus of
the chromospheric medium, greatly reducing the compression, hence the
radiative losses that act once a shock has developed.
This might explain the strong regional selectivity transient seismic
sources show favoring sunspot penumbrae, where magnetic fields are
highly inclined.
This includes the frequent occurrence of seismic emission from the
magnetic neutral lines separating opposing polarities in 
$\delta$-configuration sunspots.

On the other hand, the infusion of a plasma with a strong magnetic
field can inhibit the work capacity of such a medium in response to
impulsive heating.
Because of this, transient generation due to impulsive heating 
of a magnetic medium favors denser media.
This suggests further conditions for regional selectivity, which
might relate to the affinity of transient seismic sources to regions
from which hard X-rays are seen to emanate.

In the context of our present knowledge, the foregoing hypotheses are
highly speculative. 
However, the crucial point is that the inclusion of inclined magnetic
fields is fundamental to our understanding of the role of impulsive
atmospheric heating in transient seismic emission.

The introduction of obliquely inclined fields will complicate the HD 
simulations considerably.
However, horizontal magnetic fields can be incorporated into existing
in 1-D HD codes relatively simply.
The results of relatively cheap simulations such these can serve as a
practical basis for more realistic, but expensive, 1--3-D simulations
that incorporate oblique or stochastic magnetic fields.

Bear in mind that the purpose of this study is not to promote atmospheric
heating of any kind as ``the source'' of transient seismic emisson in
acoustically active flares.
It is rather to establish the essentiality of a realistic account
for inclined magnetic fields in simulations of this phenomenon for a
realistic assessment of its contribution to transient seismic emission.
Nor should we relax into the supposition that impulsive atmospheric
heating is by any stretch the only agent that offers the regional
selectivity shown by the helioseismic signatures.
It simply must be understood that inclined magnetic fields are fundamental
to the contribution of impulsive atmospheric heating as a realistically
prospective contributor to transient seismic emission from flares.
This mechanism should not be ruled out of consideration as a major
contributor, even possibly to sole contributor, to transient seismic
emission without this adaptation of HD simulations of this phenomenon
having been developed and applied at length.

\begin{acks}
We thank Joel Allred for sharing his most recent simulations
of impulsive thick-target heating of the chromosphere, presented at
the November-2012 RHESSI workshop.
We also greatly appreciate the insight of Valentina Zharkova at the
RHESSI workshop.
We similarly appreciate consultation with George Fisher.
We are most gratefull to Kyoko Watanabe (ISIS) for sharing Hinode
observations of the flare SOLA2011-02-15 with us.
We appreciate the insights of K. D. Leka and Graham Barnes.
We finally appreciate support of this research by contracts from the
Solar and Heliospheric Physics Program of the National Aeronautics
and Space Administration.
\end{acks}

\end{article} 

\end{document}